# Phase evolution and enhanced room temperature piezoelectric properties response of lead-free Ru doped BaTiO$_3$ ceramic

**Kumar Brajesh,[1] Sudhir Ranjan[2] and Ashish Garg[1]**

[1]Department of Sustainable Energy Engineering
[2]Department of Chemical Engineering
Indian Institute of Technology Kanpur, Kanpur-208016 (India)

## Abstract

Recent years have witnessed considerable work on the development of lead-free piezoelectric ceramic materials and their structure-property correlations. The development of piezo response is a strong function of phase evolution in these materials. In this work, we report the effect of Ru doping and consequent phase evolution on maximization of piezoelectric response of polycrystalline lead-free barium titanate, depicted as Ba(Ru$_x$Ti$_{1-x}$)O$_3$ (BRT). The samples were prepared in a narrow compositional range of $0 \leq x \leq 0.03$ using conventional solid-state reaction method. Ru doping increases leakage current of BaTiO$_3$ samples attributed to increased oxygen vacancy concentration due to substitution of Ti$^{4+}$ by Ru$^{3+}$. Detailed structural analysis reveals samples exhibiting coexistence of tetragonal (space group: *P*4*mm*) and orthorhombic (space group: *Amm*2) structured phases near the room temperature reveal relatively enhanced piezoelectric properties. The BRT sample with Ru content of 2 mol% yields a maximum longitudinal piezoelectric coefficient, *d$_{33}$* of ~269 pC/N, a high strain value of 0.16% with a large remnant polarization of ~19 μC/cm$^2$ and a coercive field of 5.8 kV/cm. We propose that the "4d" orbital of Ruthenium plays a crucial role in improving the functional properties and in decreasing the ferroelectric Curie temperature (*T$_c$*). Our work provides clues into tailoring the phase evolution for designing lead-free piezoelectric materials with enhanced piezoelectric property.

**Corresponding author:**

E-mail address: kmrbrjsh9@gmail.com

# 1. Introduction

Piezoelectric materials, which generate a voltage in response to mechanical strain (and vice versa), are useful for a variety of applications [1-3]. Since 1950's, among all known piezoelectric materials, the workhorse materials for nearly all piezoelectric device applications have been Pb(Zr,Ti)$O_3$ (lead zirconate titanate or PZT) and its derivatives. In particular, its compositions close to morphotropic phase boundary or MPB exhibit excellent ferroelectric and piezoelectric response due to coexistence of tetragonal and rhombohedral structured phases. However, in recent years, the use of PZT is marred by the concerns related to toxicity of lead which has severe environmental and health concerns. As a consequence, several countries have imposed strict restrictions on the use of lead containing materials. Therefore, there is a serious need of developing Pb-free substitutes which display at least similar if not superior piezoelectric properties. Recently, researchers have reported several Pb-free (lead-free) piezoelectric ceramics exhibiting promising piezoelectric properties, especially the (K,Na)Nb$O_3$ based pseudo-binary system and the BaTi$O_3$ based pseudo-binary system [4-7]. Among these BaTi$O_3$ is exciting compound due to its ease of synthesis and reliable properties. The dielectric and piezoelectric properties of modified and pure BaTi$O_3$ ceramics have been well documented by Jaffe [1]. The compositional modifications are of particular interest, which can induce inter-ferroelectric instability and coexistence of ferroelectric phases at room temperature. The two inter-ferroelectric transitions, tetragonal-orthorhombic and orthorhombic-rhombohedral occur at ~0 °C and -90 °C respectively in pure BaTi$O_3$. Very recently, it was demonstrated that merely 2 mol% doping of Zr, Sn, Ce and Hf in BaTi$O_3$ can remarkably increase the piezoelectric properties by stabilizing the coexistence of orthorhombic and tetragonal phases at room temperature without significantly altering the Curie point [8-10]. This offers an interesting opportunity to chemically tune the orthorhombic-tetragonal and rhombohedral-orthorhombic phase transition boundaries closer to room temperature for enhanced piezoelectric properties mimicking a MPB scenario, akin to that in PZT. We note that the ionic radii of Zr, Sn, Ce and Hf are larger than Ti ions and hence they show enhanced piezoelectric properties [11,12].

Most of the earlier studies on compositional tuning of BaTi$O_3$ have focused on incorporating "3d" elements on Ti site to tune the dielectric and ferroelectric properties. In contrast doping with "4d" elements could be interesting to examine because fundamentally speaking, "4d" electron are more delocalized than "3d" electrons due to higher principal quantum number of same angular symmetry owing to the greater distance of "4d" orbital from the nucleus compared to the "3d" orbital. It can hence be expected that "4d" dopants will influence the properties of BaTiO$_3$ more strongly than "3d" dopants. Hence, it would be interesting to examine the effect of "4d" ion substitution at the Ti site of BaTiO$_3$ in terms of structure and property evolution. A "4d" element of interest is Ru$^{4+}$ which is according to Shannon ionic radii table, has radii of 0.62 Å which is greater than that of Ti$^{+4}$ (0.605 Å) [13]. Recently, piezo response force microscopy (PFM) and hysteresis loop measurements of 0.9PbZn$_{1/3}$Nb$_{2/3}$O$_3$ – 0.1PbTiO$_3$ (PZN-0.01PT), which is a relaxer–ferroelectric solid solution with a giant piezoelectric effect, revealed that the addition of Ru decreases the ferroelectric domain size, reduces the polar fraction distributed in the pseudocubic matrix, and leads to significant ferroelectric hardening due to the immobilization of domain walls [16,17]. Since, Ru$^{+4}$ has partially filled "4d" electrons in its valence shell and it tends to lie closer to the Fermi level making the system electrically conducting [14,15]. However, enhanced electrical conductivity of Ru may interfere with the electric poling process required to make a ferroelectric system piezoelectric. However, while effect of Ru doping on BaTiO$_3$ is yet to be examined, its magnetic nature may further allow material to show multiferroism at RT.

In view of this, we have investigated Ru doped BaTiO$_3$ ceramic in a narrow composition range within the dilute concentration regime such that its conductivity does not hamper the piezoelectric properties of the resulting material. In this report, we report on the effects of Ru ("4d") doping on the phase evolution, dielectric, piezoelectric and ferroelectric properties of BaTiO$_3$ ceramics prepared using conventional solid-state reaction method. We focus on the detailed structural analysis and show a correlation between dielectric properties, grain size, piezoelectric and ferroelectric properties at the ferro- ferro transition near the room temperature and attribute them to the phase evolution

## 2. Experimental details

Ba(Ru$_x$Ti$_{1-x}$)O$_3$ (BRT) ceramics were prepared in close compositional interval of $0 \leq x \leq 0.03$ using solid state synthesis route. High purity dried powders of BaCO$_3$ (99.8% Alfa Aesar), TiO$_2$ (99.8%, Alfa Aesar), RuO$_2$ (99.95%, Alfa Aesar) were milled in planetary ball mill (P5, Fritch) using acetone medium for 10-12 hrs. at room temperature at a speed of 250 rpm. Milled powder was dried and calcined by keeping in alumina crucibles at a temperature of 1100 ºC for 4 hrs. in high temperature furnace (Lenton make). The pellets were sintered at 1200°C for 6 hrs. in the same furnace by placing the pellets on a platinum foil and covering them by a alumina triangular. Then the sintered pellets were polished by emery paper and 0.2 mm of material was removed on both sides just to check for internal defects. The defect free pellets were further characterized. X-ray powder diffraction was performed using Rikagu (Smart Lab)) with $CuK\alpha_1$ radiation (λ-1.54Å). X-ray photoelectron spectroscopy (XPS) measurements were performed to analyse chemical states of the samples using PHI 5000 (Versa Probe II, FEI Inc) and XPSPEAK 4.3 software was used to analyse the spectra. Dielectric measurements were done on Novocontrol (Alpha AN) impedance analyzer. Measurement of longitudinal piezoelectric coefficient ($d_{33}$) was carried using Piezotest PM 300 berlincourt meter by poling all samples at room temperature for 15 minutes at a field of ~1kV/mm. Polarization-electric field hysteresis loop was measured with a Precision premier II loop tracer with MTI strain measurement setup. Rietveld refinement was carried out using FULLPROF software [18].

## 3. Results

### (a) Ferroelectric and Piezoelectric properties

**Fig. 1**a and b show RT polarization electric field (*P-E*) hysteresis loops for x=0.01 and x=0.02, Ru – doped $BaTiO_3$ ceramic samples. It is observed from **Fig.1c** that with increase in the Ru doping concentration, *P-E* hysteresis loops become lossy as indicated by rounding at the higher fields, which is suggestive of an increase in sample's electrical leakage current with increasing Ru-doping concentration possibly due to increased oxygen vacancy concentration. It is well known that in $BaTiO_3$, oxygen vacancies associated to $Ti^{3+}$ ions are present at Ti site in abundant quantities which provides larger concentration of free electrons/carriers that contribute to leakage current. On doping Ruthenium in lattice of $BaTiO_3$, $Ru^{3+}$ substitutes $Ti^{4+}$ and $Ti^{3+}$ which requires oxygen vacancy for charge neutrality according to the following defect reaction:

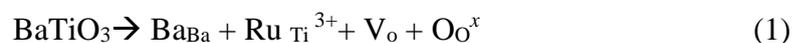

$$BaTiO_3 \rightarrow Ba_{Ba} + Ru_{Ti}^{3+} + V_o + O_o^x \qquad (1)$$

XPS analysis of the doped samples was performed to further analyze the spectra of O1s and Ti2p core level fitted by Lorentzian-Gaussian functions as shown in **Fig.2**. The high resolution XPS peak of O1s core level **(Fig 2(d-f))** for the samples were deconvoluted into two peaks at ~528.6 eV corresponding to lattice oxygen ($O_L$) and at ~531.3 eV assigned to surface oxygen vacancies defects or surface adsorbed oxygen ($O_v$). It is noteworthy that higher energy peak can be also due to surface contamination or adsorbed oxygen but since doped samples were prepared under similar conditions, hence no significant difference due to surface contamination is expected. On increasing doping concentration of Ru (*x* = 0.010, 0.020 and 0.025), relative area of $O_v$ peak [$O_v$/ ($O_v$_+ $O_L$)] gradually increased from 0.58 (*x* =0.010) to 0.87(*x* =0.025), which demonstrates the increase in concentration of oxygen vacancies by substitution of $Ti^{4+}$ by $Ru^{3+}$ leading to enhanced leakage current with Ru doping concentration. Moreover, during preparation of $BaTiO_3$ few $Ti^{4+}$ ions are converted to $Ti^{3+}$ whose concentration is further enhanced upon Ru doping which can be expressed by

$$Ti^{4+} + e' \rightarrow Ti^{3+} \quad (2)$$

This $Ti^{3+}$ together with lattice oxygen results in creation of more oxygen vacancy concentration which can be illustrated by the following steps:

$$2Ti^{3+} + 3O^{2-} \rightarrow 2\,(Ti_{Ti(4+)})' + 2h^{\cdot} + 3O^{x}_{O} \quad (3)$$

$$2O^{2-} + 4h^{\cdot} \rightarrow O_2 \uparrow + V_O \quad (4)$$

The fitted Ti2p XPS spectra of the Ru doped samples are shown the **Fig. 2(a-c)**, where ratio of $Ti^{3+}/Ti^{4+}$ gradually increases from 0.22 for $x = 0.01$, 0.28 ($x = 0.020$) to 0.36 for $x = 0.025$ samples, which further confirms that $Ti^{3+}$ concentration in $BaTiO_3$ lattice gradually increases on Ru doping. Therefore, Ru doping increases the concentration of ionized oxygen vacancy, free electron and the hole resulting in increased leakage current with Ru doping. Further the rate of increase in the current with field also increases with doping. This can be attributed to the electronic states of "$Ru^{4+}$" which lie near the fermi level, thus making the system electrically more conducting in nature which provides pathways for the movement of free charges causing electrical leakage. Because of large leakage current at higher Ru doping, the ferroelectric hysteresis loop does not appear after x=0.02. **Fig. 1d** shows the composition variation and longitudinal piezoelectric response ($d_{33}$) of BRT in the composition range of $0 \leq x \leq 0.03$. The longitudinal piezoelectric coefficient ($d_{33}$) increases from 190 pC/N for pure $BaTiO_3$ to 242 pC/N at $x=0.01$, 254 pC/N at $x = 0.015$, and then reaching a maximum of 269 pC/N at $x = 0.02$ after reducing to 261 pC/N at $x=0.025$ and 232 pC/N at x=0.03. Latter is because of inability of $RuO_2$ to further solubilize in $BaTiO_3$ powder after $x = 0.02$ as also depicted from X-ray diffraction patterns shown in subsequent sections. The piezoelectric and ferroelectric properties of polycrystalline ceramics also depend on grain size and as a general rule, piezoceramics with smaller grain size exhibits lower polarization and $d_{33}$ [19]. To examine this, we studied the microstructure of the samples using scanning electron microscopy (SEM) and **Fig. 3**(a-d). shows the SEM images of BRT ceramics with varying ruthenium content ($x$) from 0.01 to 0.025. The images show that the grain size of Ru-doped BT ceramics increases with increasing Ru content and depicts a similar trend as shown by the piezoelectric constant ($d_{33}$) values. Further, we measured the bipolar strain of the samples as a function of Ru doping. **Fig. 4**(a-d) show the variation of bipolar strain of Ba(Ti$_{1-x}$Ru$_x$)O$_3$ ceramics at $x = 0$, $x = 0.01$, $x = 0.02$ and $x = 0.025$, measured at room temperature. The figures show a bipolar strain of 0.11% for $x = 0.01$ and 0.16% for $x = 0.02$, which is higher than that obtained for pure BaTiO$_3$ (strain ~ 0.1%), achieved at a field strength of ~38 kV/cm. This is a reasonably good value for a lead-free ceramics system, which can be compared with lead containing ceramics such as Pb(Zr,Ti)O$_3$ ceramics ($d_{33}$ ~500 pC/N) and electro strictive ceramics Pb(Mg,Nb)O$_3$-PbTiO$_3$ [20]. These results emphasize that one can obtain a large strain of ~ 0.16% in dilute 'Ru' modified BaTiO$_3$ ceramics with Ru content of 2.0 mol% making this system a potential candidate for replacing lead based piezoelectric ceramics for device applications.

*(b) Temperature dependent dielectric constant and X-ray diffraction studies.*

Further, we measured the changes in the dielectric constant of BRT samples as a function of Ru doping over the temperature range from room temperature to 200°C at a frequency 1 kHz and the results are shown in **Fig. 5** which reveals that as Ru content is increased (within the solubility limit), the Curie temperature ($T_c$) shifts to a lower temperature. The $T_c$ of the samples doped with more than 2.0 mol. % Ru is independent of Ru content. This indicates that the solubility limit of Ru is approximately 2.0 mol%. The dielectric data also suggests that as compared to the three phase transitions in pure BaTiO$_3$ which occur at $T_c$ (the paraelectric to ferroelectric phase transition) = 120 °C, $T_2$ = 0 °C, and $T_3$ = -90 °C respectively, $T_c$ for BRT crystals occurs at lower temperatures with increasing Ru concentration. The decrease in $T_c$ with Ru concentration is also confirmed by high temperature XRD analysis as shown in **Fig. 6** which exhibits that the $T_c$ for $x = 0.01$ is 120 °C, decreasing further to 110 °C for $x = 0.015$ and 90 °C for $x = 0.02$ as evident from the structural transition from tetragonal phase (depicted by 200/220 doublet) to cubic phase (depicted by 200/222 singlet). We will discuss the structural evolution in detail in the subsequent section. Further, **Fig. 7** shows the temperature dependent real part and imaginary parts of permittivity for $x = 0.01$ sample at a frequency of 1 kHz. The figure shows two phase transitions: cubic-tetragonal phase transition at ~110 °C as depicted by a sharp peak and a tetragonal-orthorhombic phase transition occurring near room temperature (RT~30 °C) as depicted by small step in Tan $\delta$ *vs* Temperature plot, as shown in **Fig. 7** inset.

*(c) Detailed structural analysis of Ba(Ti$_{1-x}$Ru$_x$)O$_3$ ceramics*

To further examine the structure evolution as function of Ru doping in BaTiO$_3$, we first investigated the Bragg profiles of BRT, as shown in **Fig.** 8 using XRD patterns obtained at RT, measured as a function of composition. While a visual inspection of the diffraction patterns seems to suggest that the compositions with $x = 0.01$ to 0.03 show presence of only tetragonal phase, as evident from {111}$_{pc}$ singlet and doublet of {200}$_{pc}$ (pc denotes pseudocubic), a careful analysis reveals that the distinct change in profiles shape becomes noticeable, and the intensity of {001}$_{pc}$ decreases on increasing the Ru concentration. In contrast, profiles of peaks at higher angles such as {222}$_{pc}$ exhibit additional features such as flattening (or smoothening) of the peak and appearance of intermediate humps, which are marked by arrows in **Fig. 7**a. The appearance of humps clearly indicates the possibility of an additional phase in this system at RT. For confirmation of the coexistence of another phase along with the tetragonal phase at RT in BRT, we conducted Rietveld refinement of the spectra involving minimization of overall residual between the calculated and the observed counts. Strong peaks in the pattern were given more statistical significance in the procedure and hence fitting parameters are biased by the best fitting of strong peaks in the pattern. Since, in Ru doped BT system i.e. BRT, additional peaks intensity is very low, therefore distortion of tetragonal structure is also likely to be very small. A combination of two space groups *P4mm* and *Amm*2 was considered for the refinement. This is because BaTiO$_3$ undergoes tetragonal (*P4mm*) to orthogonal (*Amm*2) phase transition at ~ 0 °C, and that the orthorhombic phase may have appeared as a precursor well before the actual thermodynamic phase transition temperature of 0 °C [21]. The initial value of orthorhombic lattice parameters of the *Amm*2 phase were kept at $a_o = c_T$, $b_o = c_o = \sqrt{2}a_T$, as its lattice parameters closely resemble those of tetragonal phase (the subscripts O and T depict the orthorhombic and tetragonal phases respectively). The first step in the fitting strategy involved fitting the pattern with a single phase tetragonal with space group *P4mm* model to accurately account for tetragonal peak positions. In the subsequent step, second phase with orthorhombic structure with space group *Amm*2 was added and its lattice parameters were refined while keeping the parameters of the *P4mm* phase fixed. Since, starting lattice parameters of the orthorhombic phase predicted peak position close to the tetragonal peak, the program was expected to find suitable lattice parameter to account for the additional orthorhombic peaks. At last, both lattice parameters were refined. The *Amm*2+*P4mm* model yielded the best "goodness of fit" with $\chi^2$ = 1.21 for x = 0.01, $\chi^2$ = 1.26 for x = 0.02 and $\chi^2$ = 3.68 for $x$ = 0.025 respectively, whereas only *P4mm* (Tetragonal) model for $x$ = 0.01 yielded the worst fit, pointed by arrows as shown in **Fig. 9**a. **Fig. 9**(a-d) shows the Rietveld fitted powder XRD patterns of the three representative compositions $x$ = 0.01, 0, 0.02 and 0.025 respectively, with all the refined parameters shown in **Table 1**. Goodness of fit becomes worse for $x$ = 0.025 and this is attributed to the appearance of a few peaks corresponding to secondary non-perovskite phases of $RuO_2$ which conformed by JCPDS file no. 002-1365 as visible in the profile shape after $x$ = 0.02, marked with star (*) in **Fig. 8**(d-e). Appearance of these secondary phases indicates that the solubility limit of $RuO_2$ in $BaTiO_3$ powder is $x$ = 0.02.

## 4. Discussion

The detailed structural analysis clearly reveals that the composition range 0.01< x <0.03, which exhibits enhanced piezoelectric coefficient is characterized by coexistence of two ferroelectric phases (tetragonal + orthorhombic) at room temperature. A correlation is found at morphotropic phase boundary (MPB) in lead-based material such as PZT, separating tetragonal and rhombohedral phase regions [1] and that the phase coexistence at the MPB leads to maximization of properties. However, there are some distinct differences between the nature of the MPB region found in lead-based materials and PPT (polymorphic phase transition) present in the lead-free materials. The lead based MPB materials exhibit low symmetry monoclinic phase (space group *Pm*) with pseudocubic direction $(101)_C$ which allows the availability of low energy continuous polarization rotation pathway(s), and, thereby, contributes significantly to the intrinsic piezo-response of the system [22 -25]. Alternatively, in the framework of adapting phase theory, the formation of nano-domains also contributes significantly to the overall piezoelectric response in ferroelectrics [10]. In the present case, the polarization vectors in the tetragonal and orthorhombic phase are along $[001]_c$ and $[101]_c$ pseudocubic direction respectively. The plane $(101)_c$ provides a continuous pathway for the polarization vector to rotate from $[001]_c$ towards $[101]_c$ direction. This can happen in the two-phase model consisting of *P4mm* and *Amm*2 phases as found in the present system BRT without necessitating the need for a low symmetry monoclinic phase for any of the compositions. Another evidence of ferroelectric-ferroelectric coexistence near the room temperature is found in temperature dependent real and imaginary parts of dielectric constants as shown in the inset of **Fig.7**. The figure shows a direct correlation between the powder XRD patterns and dielectric spectra. **Fig. 10.** shows the plot of the unit cell volume of tetragonal with various composition. The decrease in the cell volume of tetragonal structure may be due to difference in the electronic configuration of "$Ti^{4+}$" and "$Ru^{4+}$" ions. The ferroelectric tetragonal phase has a larger cell volume than the paraelectric cubic phase and hence a spontaneous tensile strain develops in the tetragonal phase to create the ferroelectric state. Further, the cell volume of the tetragonal phase decreases with an increase in Ru content which indicates that the intrinsic strain is compressive instead of tensile thereby contradicting the effect. The decrease in cell volume of Ru- doped samples does not permit Ti(Ru) off–centering due to less space in the unit cell. In $BaTiO_3$, Ti $3d°$ bands are completely empty due to charge transfer. Recently B. Sarkar et al reported that the electronic configuration of Ru plays a crucial role in determining the value of $T_C$ rather than the size of doped $BaTiO_3$ system. Therefore, the decrease in $T_c$ causes a decrease in the unit cell volume of the tetragonal phase associated with ferroelectric order [26]. Overall, our results unambiguously demonstrate that the enhanced piezoelectric response in the closed composition intervals in the dilute concentration regime is associated with the preservation of the phase coexistence (*P4mm+Amm2*) state.

## 5. Conclusions

In conclusion, a comparative study of structure and piezoelectric properties was performed for Ru doped $BaTiO_3$ ceramics synthesized through solid-state reaction route, with sintering at 1200°C for 6 hrs. XPS results confirm the Ru doping in $BaTiO_3$ by increasing the oxygen vacancy concentration (or $Ti3^+$ concentration) which results in increased leakage current and makes doped samples more conducting. The detailed structural analysis clearly reveals that there exists a narrow composition range of Ru doping within which samples exhibit relatively enhanced piezoelectric coefficient which is characterized by coexistence of two ferroelectric phases (*P4mm+Amm2*) appearing near the room temperature. The longitudinal piezoelectric coefficient $d_{33}$ value of the poled sample was found to be ca. 269 pC/N with high strain value of 0.16% and a substantial remnant polarization of 19.24 μC/cm$^2$ with a coercive field of 5.8 kV/cm. Cell volume of the tetragonal phase decreases with an increase in Ru content indicating that the intrinsic strain is compressive instead of tensile which contradicts the size effect. From this view, a large strain of ~0.16% in dilute 'Ru' modified $BaTiO_3$ makes this system a strong Pb-free contender for actuator applications. Our work may provide new strategies for designing Pb-free piezoelectric materials with enhanced piezoelectric property by stabilizing phases of certain types.

**Conflict of Interest**

All participating authors declare no financial or other type of conflict of interest.

**Acknowledgements**

Authors thank the Science and Engineering Research Board (SERB), Government of India, for the financial support through grant no EMR/2017/003486.

# Figures Caption

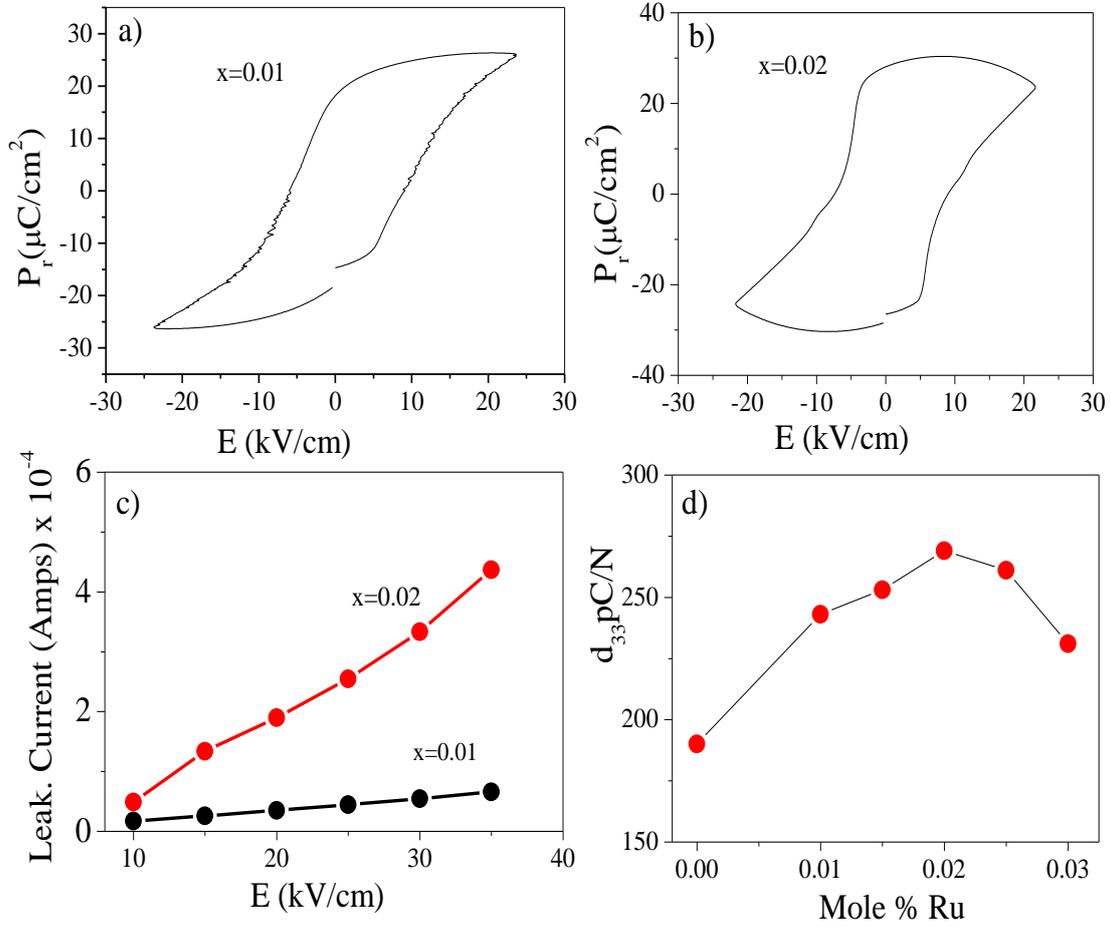

**Figure 1.** Polarization and leakage current versus electric field of the Ba(Ti$_{1-x}$ Ru$_x$)O$_3$ ceramics at (a) x=0.01 and (b) x=0.02 measured at room temperature, and (d) Piezoelectric coefficient of the Ba(Ti$_{1-x}$ Ru$_x$)O$_3$ ceramics as a function of x.

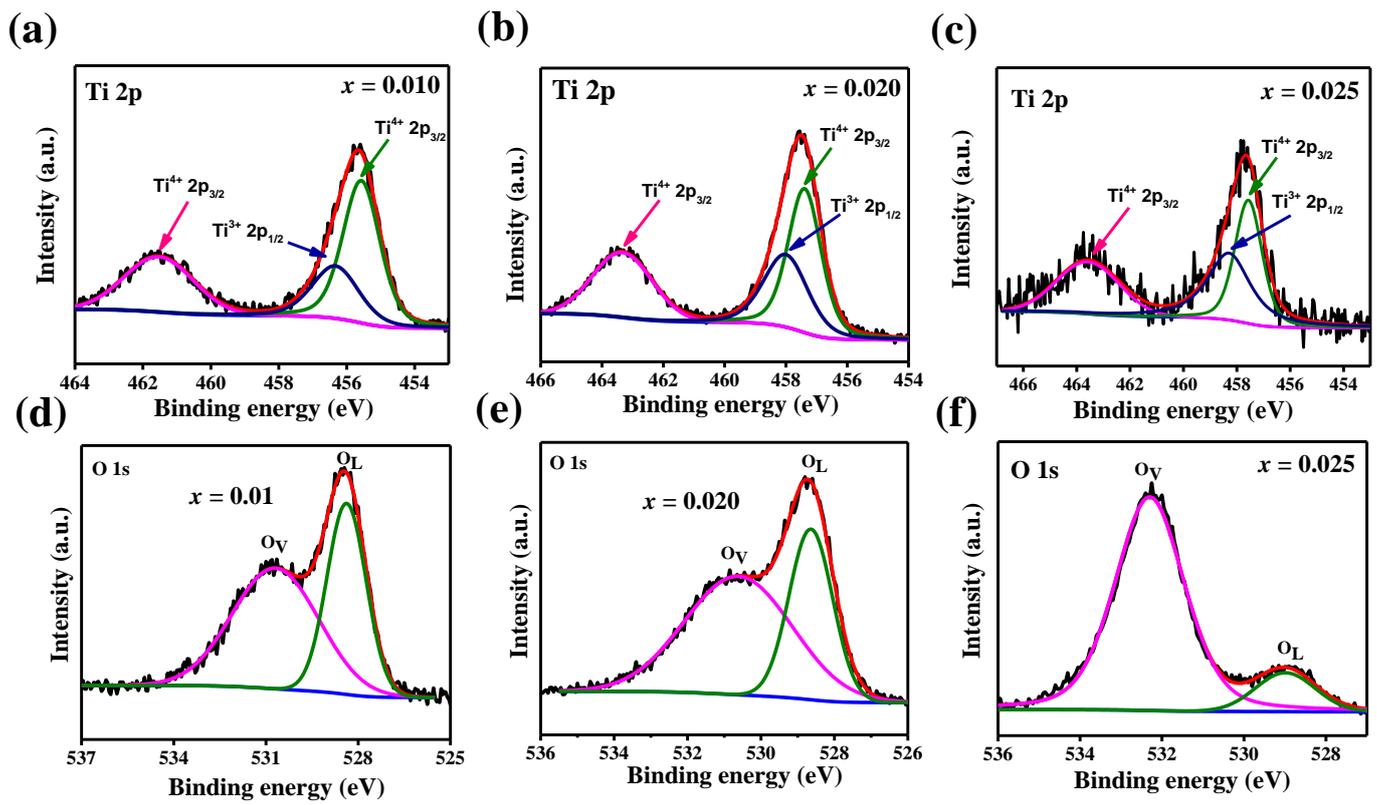

**Figure 2.** High resolution XPS spectra of Ti 2p and O1s for (a,d) *x*=0.01, (b,e) *x* = 0.02 and (c,f) *x* = 0.025 samples, respectively.

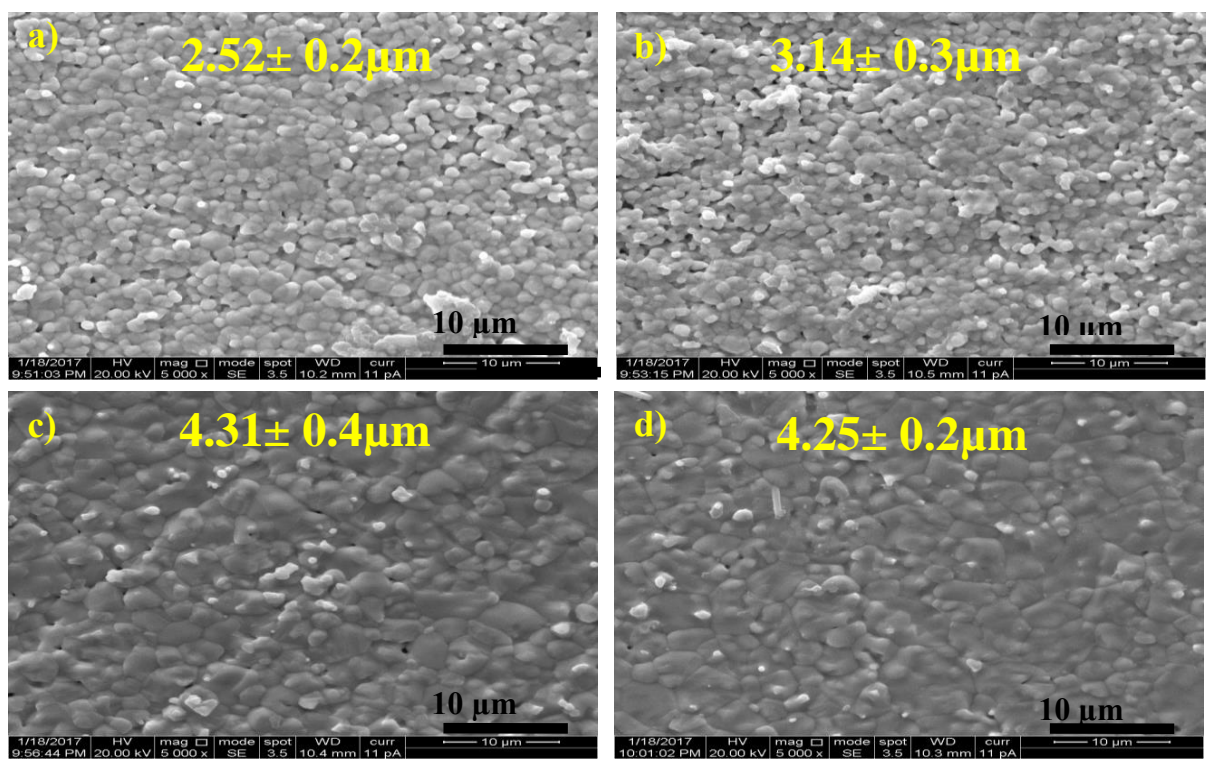

**Figure 3** (a-d) shows the microstructural images with average grain size of BRT ceramics with various ruthenium content (x) of (a) = 0.01, (b) =0.015, (c) =0.02 and (d) = 0.025 at room

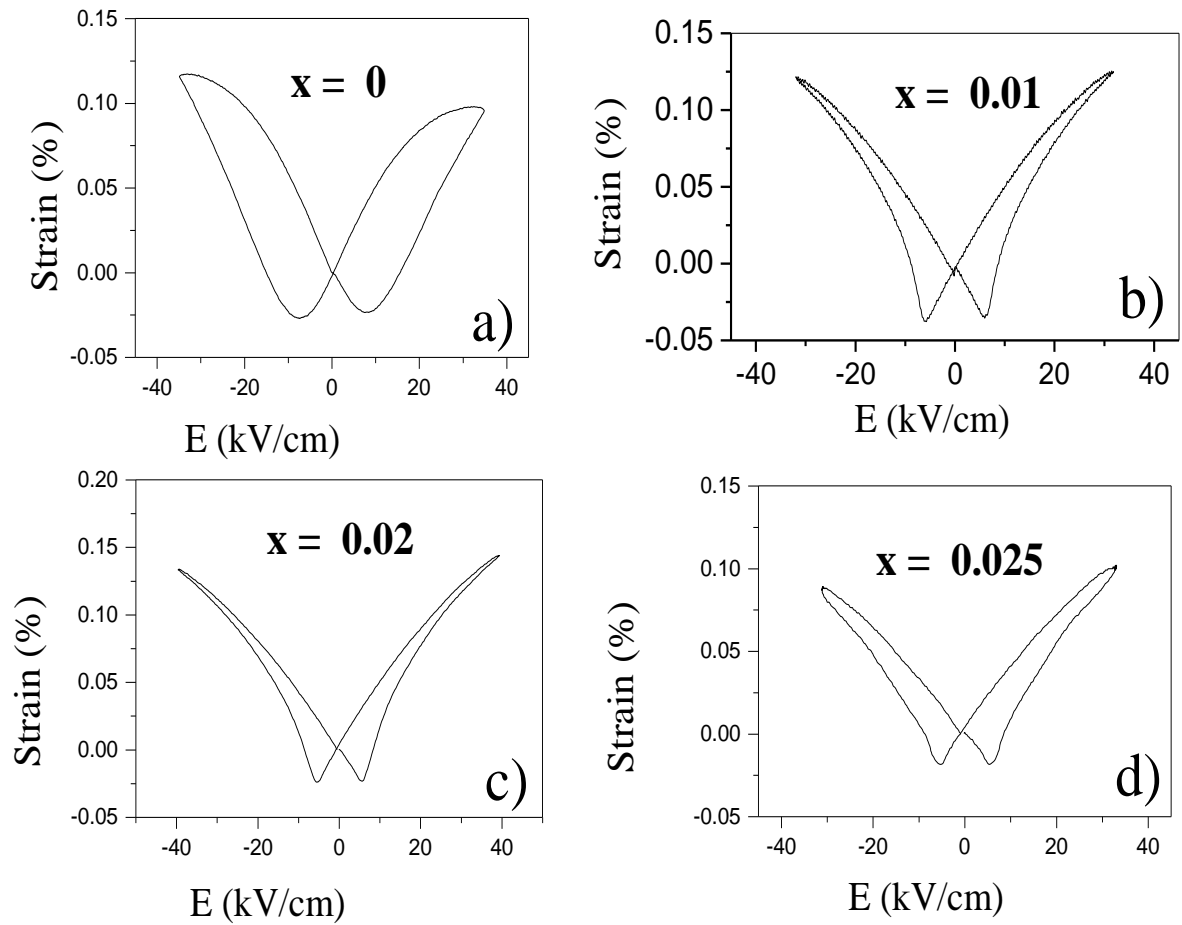

**Figure 4.** Electric field induced bipolar strain for the Ba(Ti$_{1-x}$ Ru$_x$)O$_3$ ceramics at (a) x=0, (b) x=0.01, (c) x= 0.02 and (d) x= 0.025 measured at room temperature.

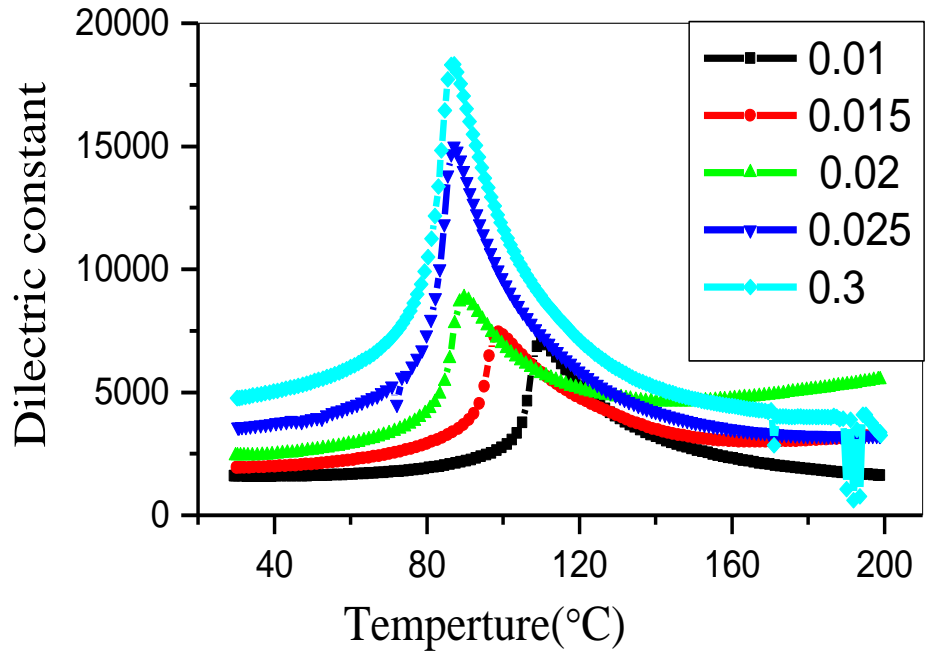

**Figure 5**. Temperature dependence of dielectric constant for the Ba(Ti$_{1-x}$ Ru$_x$)O$_3$ ceramics at x=0.01, x=0.015, x=0.02, x= 0.025 and x= 0.03 measured at 1kHz.

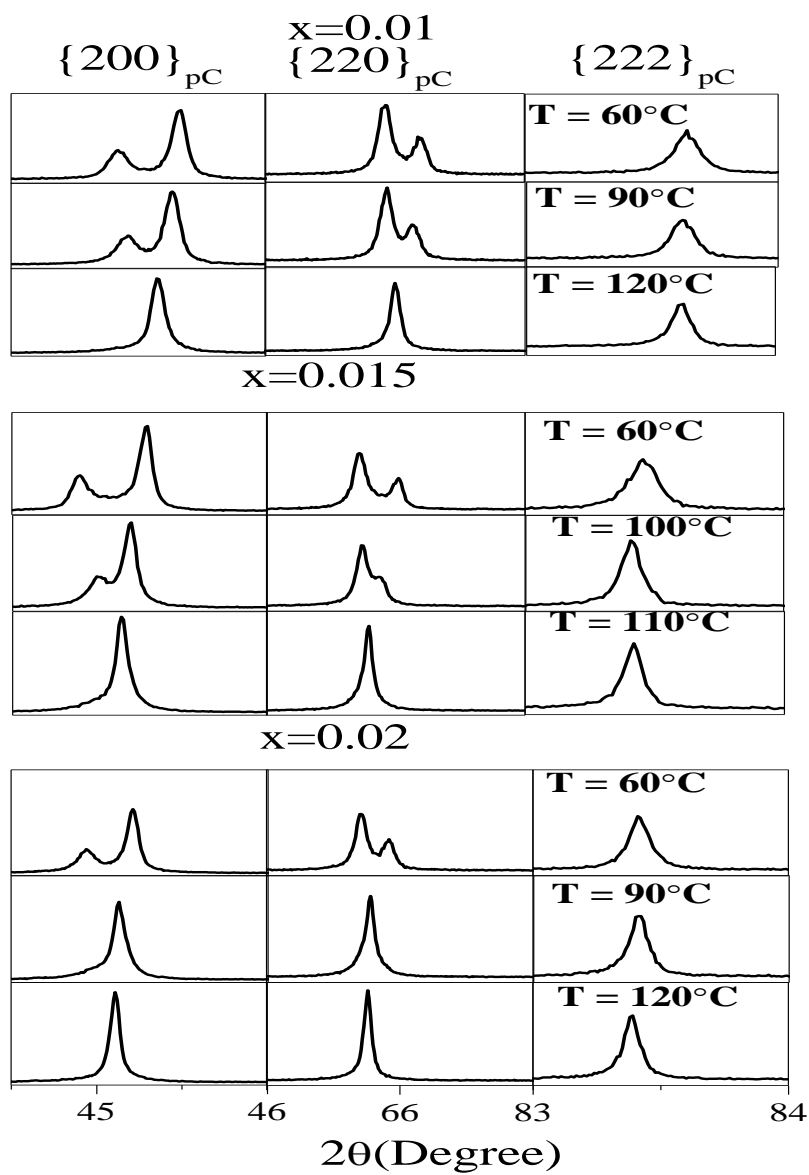

**Figure 6.** Evolution of pseudocubic {200}, (220) and {222} X-ray powder diffraction Bragg profiles of Ba $(Ti_{1-x} Ru_x)O_3$ with composition at high temperatures.

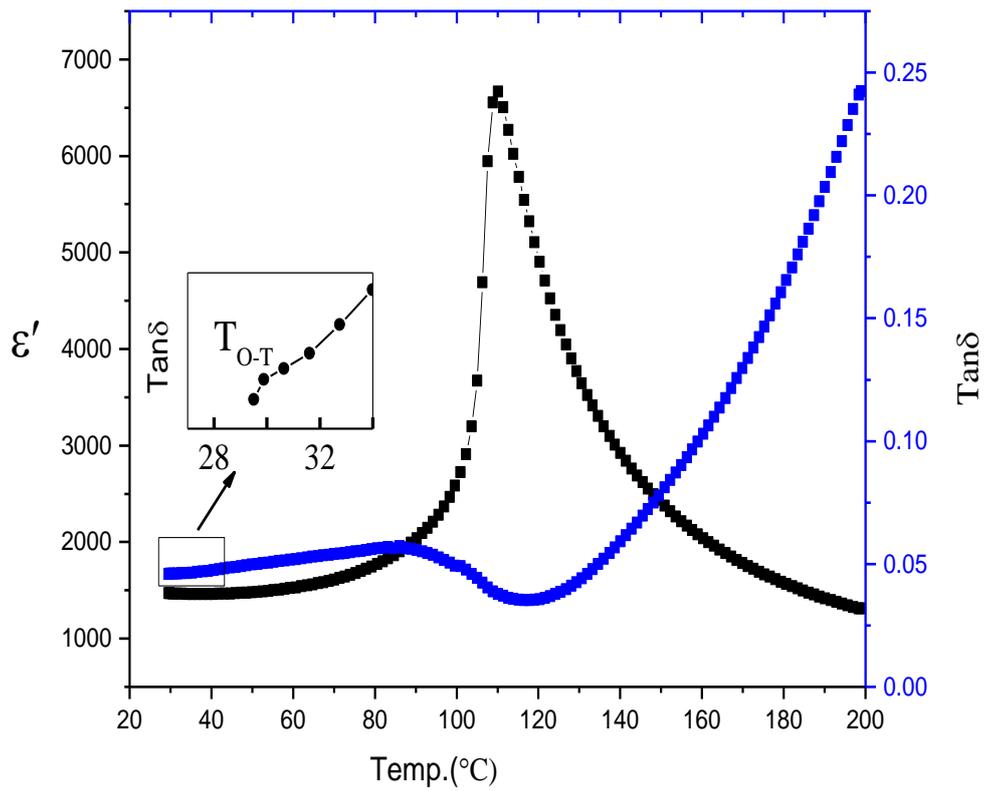

**Figure 7.** Temperature dependence of real part and imaginary part of Ba(Ti$_{1-x}$ Ru$_x$)O$_3$ ceramics for x=0.01 measured at 1Khz.

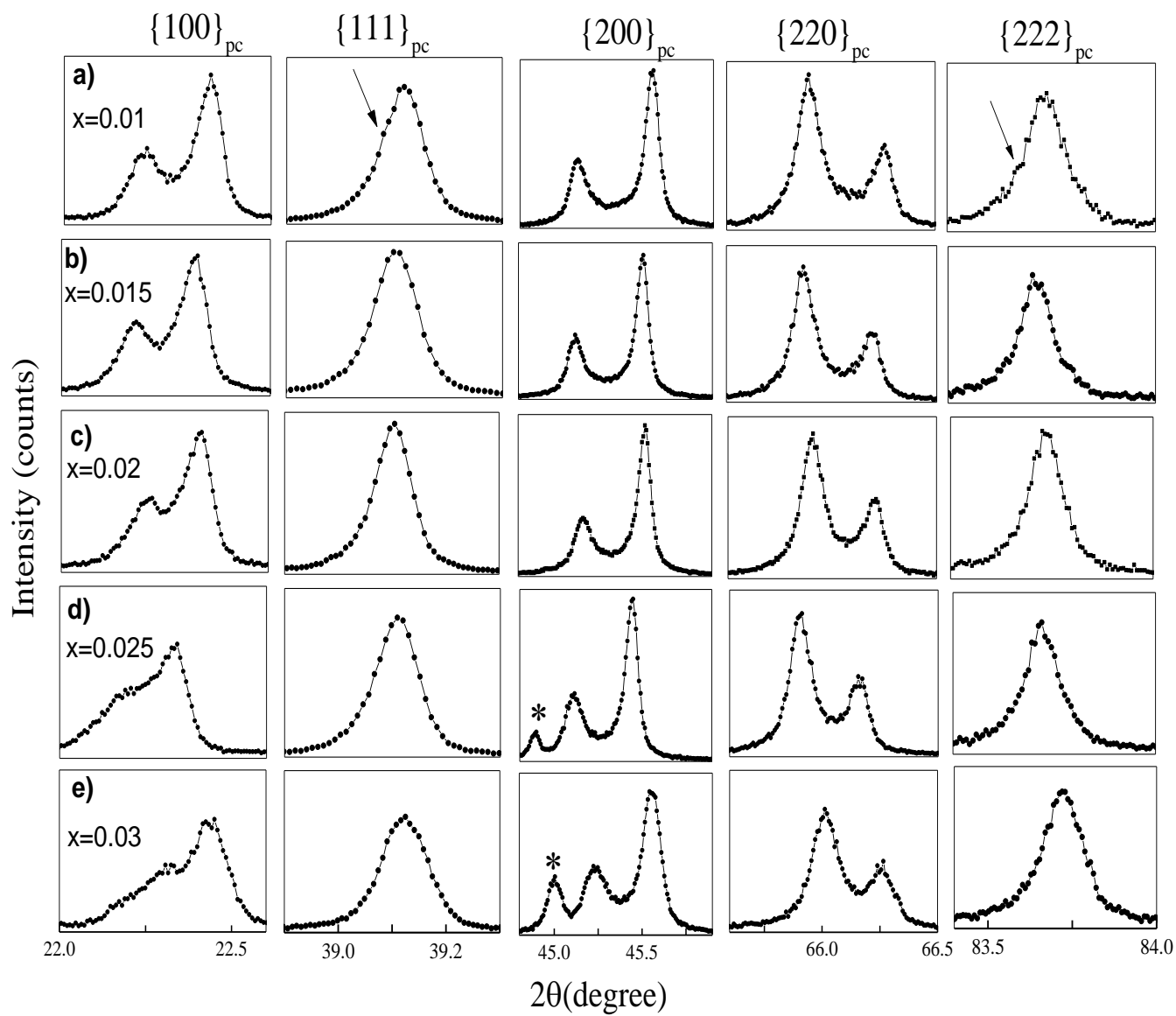

**Figure 8**. Evolution of pseudocubic {100},{111},( {200},(220) and {222} ) X-ray powder diffraction Bragg profiles of Ba(Ti$_{1-x}$ Ru$_x$)O$_3$ with composition at room temperature. The stars (d & e) highlight the impurity phase in the profile.

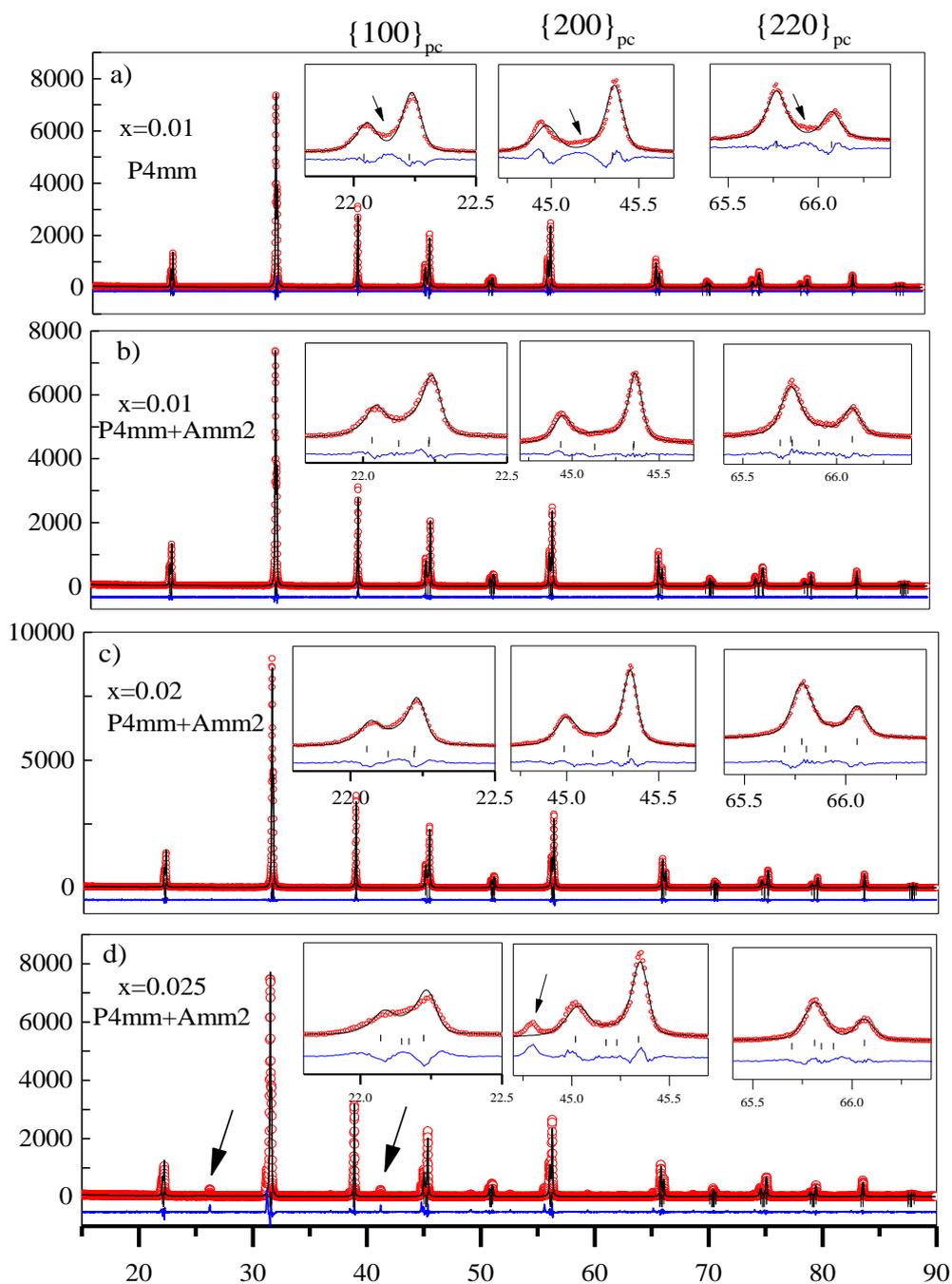

**Figure 9.** Rietveld fitted x-ray powder diffraction of XRD profiles of Ba Ti$_{1-x}$Ru$_x$O$_3$ for (a-b) x=0.01, (c) x=0.02 and (d) x = 0.025. The insets in the three plots shows magnified Bragg profiles corresponding to the pseudo-cubic {100} (left inset), {200} (middle inset) and {220}(right inset). An arrow shows the impurity phase whereas the vertical bars shows the Bragg peak position of the perovskite phase.

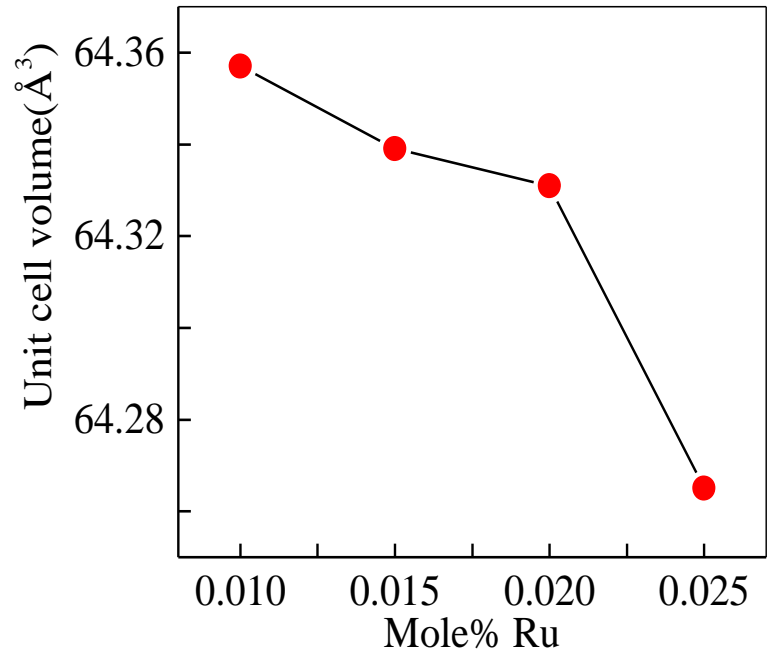

**Figure 10.** Unit cell volume of tetragonal with various composition.

Table I: X-ray powder diffraction refined structural parameters and agreement factors for Ba(Ti$_{0.99}$Ru$_{0.01}$)O$_3$ using Tetragonal(*P4mm*) + orthorhombic(*Amm*2) phase coexistence models.

| | Space group: P4mm | | | | Space group: Amm2 | | | |
|---|---|---|---|---|---|---|---|---|
| Atoms | x | y | Z | B(Å$^2$) | x | y | z | B(Å$^2$) |
| Ba | 0.000 | 0.000 | 0.000 | 0.16(2) | 0.000 | 0.000 | 0.000 | 0.25(2) |
| Ti/Ru | 0.500 | 0.5000 | 0.480(1) | 0.13(1) | 0.500 | 0.000 | 0.483(1) | 0.24(3) |
| O1 | 0.500 | 0.5000 | -0.006(2) | 0.11(0) | 0.000 | 0.000 | 0.497(3) | 0.12(0) |
| O2 | 0.500 | 0.000 | 0.516(2) | 0.12(2) | 0.500 | 0.267(2) | 0.268(0) | 0.42(1) |
| a=3.9956 (4) Å, c= 4.0311 (5) Å   v= 64.35 (1) Å$^3$ , %Phase = 74.37(2) | | | | | a= 3.995(3) Å, b= 5.675(4) Å, c= 5.679(0) Å   v= 128.80 (1) Å$^3$, %Phase = 25.63(3) | | | |
| R$_p$: 8.90,   R$_{wp}$: 13.2,   R$_{exp}$: 11.85,  Chi$^2$: 1.24 | | | | | | | | |